\documentclass[reprint,amsmath,amssymb,aps]{revtex4-1}

\usepackage{graphicx,dcolumn,bm}
\usepackage{epsfig}

\begin{document}

{\bf Comment on ''Narrow Structure in the Excitation Function of \boldmath$\eta$\
photoproduction off the Neutron"}\\[1ex]
A.V.~Anisovich\hspace{0.5mm}\mbox{$^{1,2}$},
E.~Klempt\hspace{0.5mm}\mbox{$^{1}$},
V.~Nikonov\hspace{0.5mm}\mbox{$^{1,2}$},\\
A.~Sarantsev\hspace{0.5mm}\mbox{$^{1,2}$}, and
U.~Thoma\hspace{0.5mm}\mbox{$^{1}$}
\\[1ex]
$^{1}$ HISKP, Universit\"at Bonn, Germany\\
$^{2}$ St. Petersburg NPI, Gatchina, Russia \\[1ex]
In a recent letter, observation of a narrow structure was reported in
photoproduction of $\eta$ mesons off neutrons bound in $^2$H and $^3$He
\cite{Werthmuller:2013rba}. The data from both data sets agree well,
hence we restrict the discussion to the results using deuterons. The
structure had been observed before
\cite{Kuznetsov:2006kt,Jaegle:2008ux} and is listed in the Review of
Particle Properties as one--star resonance \cite{Beringer:1900zz}. The
new data exceed the earlier data both in quality and in statistics. In
the new experiment, the hit proton or neutron was detected, and hence
the $N\eta$ invariant mass was reconstructed without smearing due to
the Fermi motion. These experimental achievements greatly enhanced the
visibility of the narrow structure. The contributions from $\eta$
production were determined by two different methods; the
differences were used to estimate the systematic error. In the analysis
below, we added both errors (quadratically).

In \cite{Werthmuller:2013rba}, the structure was tentatively
interpreted as a narrow nucleon resonance at $W=(1670\pm5)$\,MeV mass
and $\Gamma=(30\pm5)$\,MeV width. The product coupling of the
hypothesized resonance given by its helicity amplitude $A_{1/2}^n$ and
the branching ratio for its neutron-$\eta$ decay $b_\eta$ was
determined to
$\sqrt{b_\eta}A_{1/2}^n=(12.3\pm0.8)\,10^{-3}$\,GeV$^{-1/2}$. The
structure is exciting since a $P_{11}$ resonance with precisely these properties
is predicted as the non-strange member  of an anti-decuplet
\cite{Polyakov:2003dx}.

The authors of \cite{Werthmuller:2013rba} admitted that the older less
precise data could be described without introducing a new nucleon
resonance with exotic properties. In this comment we confirm that also
the new and precise data are well described by the interference within
the $S_{11}$ wave. Moreover, the data are incompatible with the
existence of a nucleon resonance with the reported properties.

The data were fitted within the Bonn-Gatchina partial-wave analysis.
Masses, widths, decay couplings of resonances were all frozen by a fit
to $\pi N$ elastic and inelastic reactions and to data on
photoproduction off protons (see
\cite{Anisovich:2011fc,Anisovich:2013vpa} for the data included). Here,
only the helicity amplitudes for photoproduction of nucleon resonances
off neutrons and contributions from $t$- and $u$-channel exchange were
used as free parameters. To constrain the data on $\gamma n\to \eta n$
further, we included GRAAL data on the beam asymmetry for this reaction 
and data on $\gamma n\to \pi N$. These additional data were described well 
and have no impact on the conclusions; hence we retain from a more detailed
discussion of those data.

\begin{figure}[ph]
\centerline{\epsfig{file=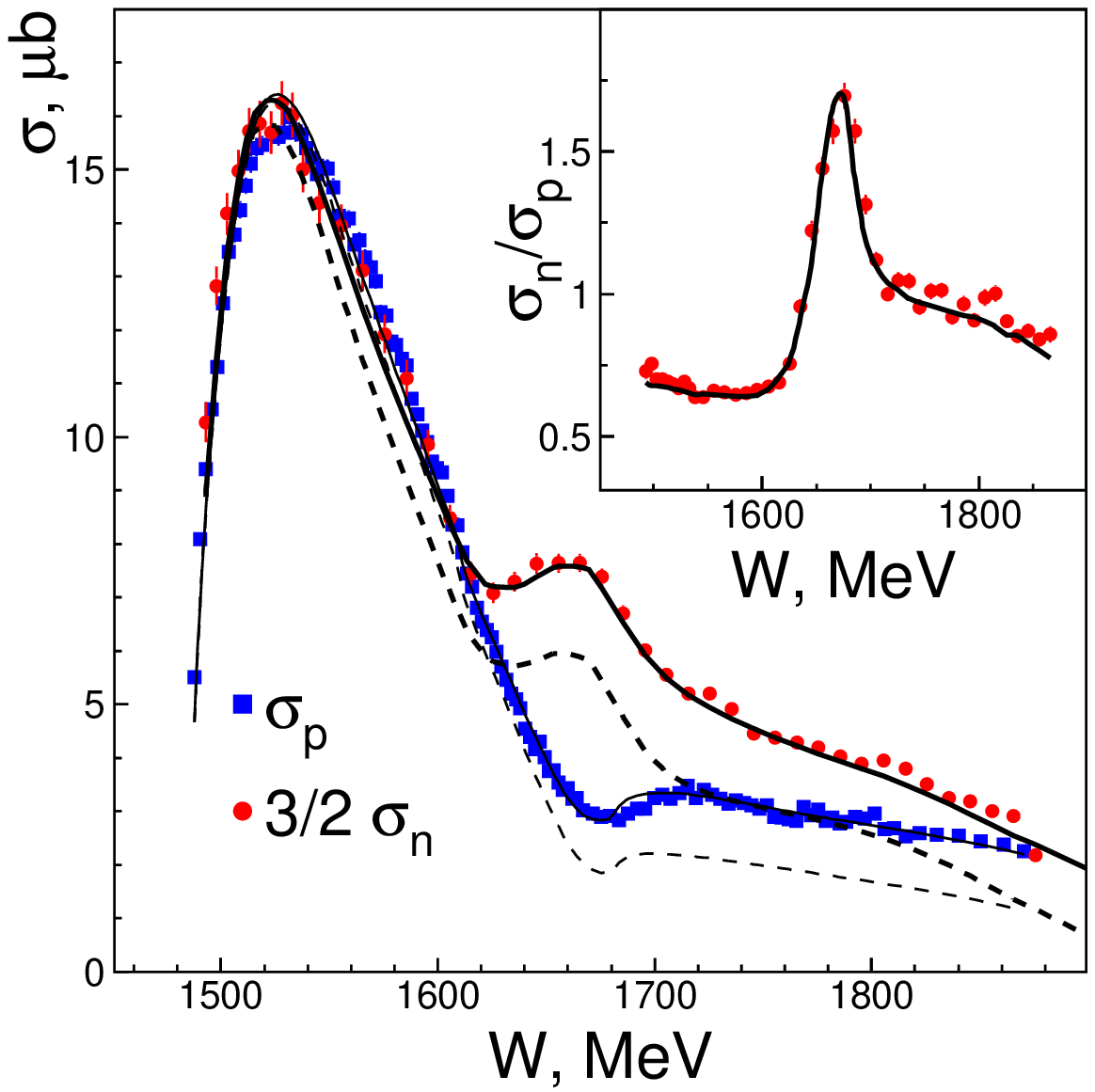,width=0.24\textwidth}
            \epsfig{file=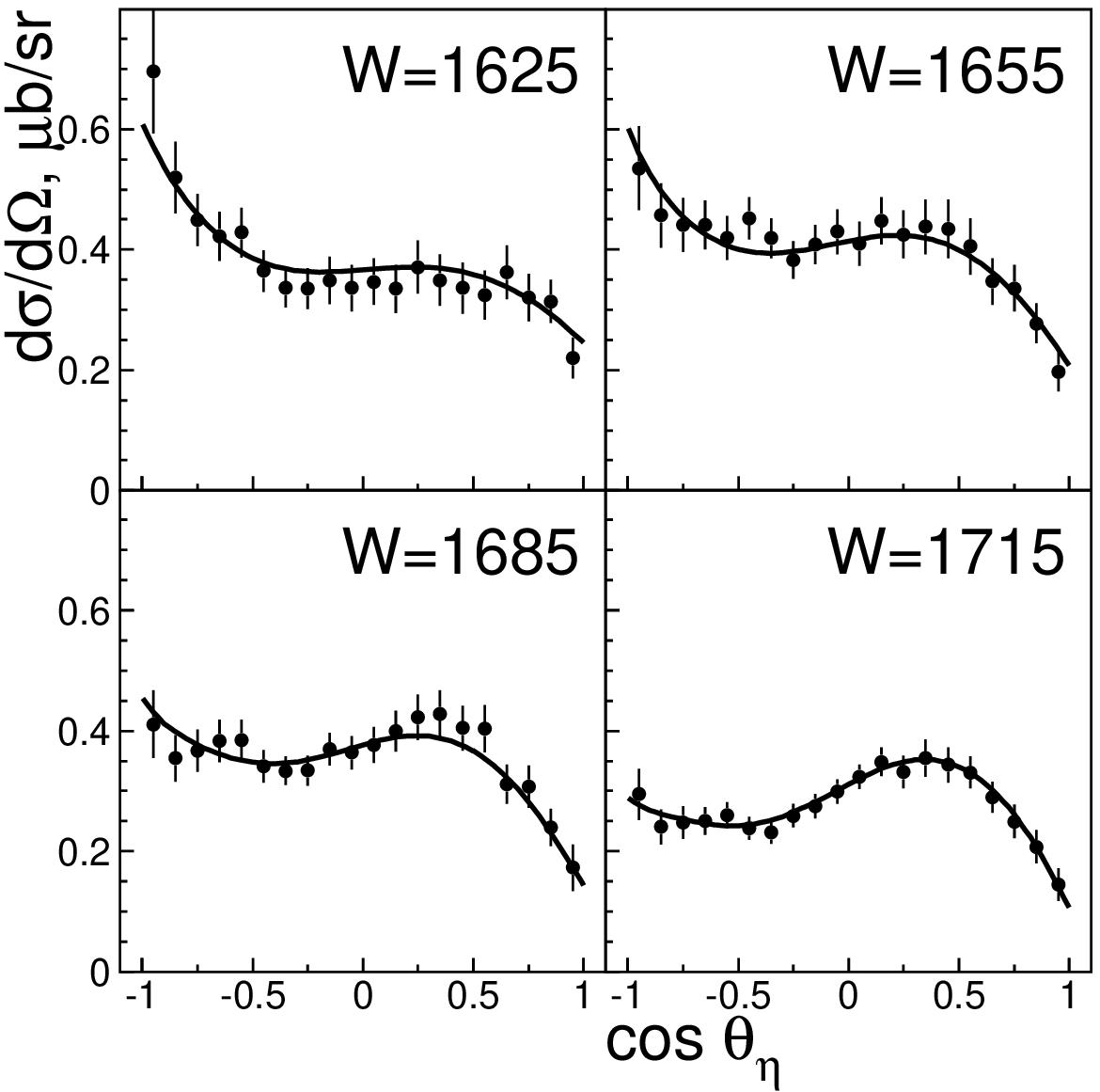,width=0.24\textwidth}}
\caption{\label{tot}(Color online) Left: The total cross section for
$\gamma n\to \eta n$ (multiplied by 3/2), $\gamma p\to \eta p$, and
their ratio (as inset). The solid curves represent our fit folded with the experimental resolution (thick $\eta
n$,  thin $\eta p$), the dashed curves the contributions from the
$S_{11}$ waves. Right: Selected differential cross section for $\gamma n\to \eta
n$ in the region of the narrow structure.}
\end{figure}

Fig.~\ref{tot} shows the total cross section for $\gamma p\to \eta p$, for  $\gamma n\to \eta n$ and - as insert - the ratio $\sigma_n/\sigma_p$, and our fit to the data. Overall, the description of all angular distributions is excellent, with $\chi^2=95$ for 200 data points in the $(1610 - 1710)$\,MeV mass range  (or 620 if statical errors only are used). If a resonance with properties from \cite{Werthmuller:2013rba} is enforced, the $\chi^2$ increases by 257
units. Just acceptable solutions -- with an increase in $\chi^2$ of about 25 -- are obtained for
$(-3<\sqrt{b_\eta}A_{1/2}^n<5)\cdot 10^{-3}$\,GeV$^{-1/2}$. 

The reason for the peak structure in $\eta n$ and dip structure in $\eta p$ lies in the opposite relative sign of the helicity amplitudes for the two resonances $N(1535)S_{11}$ and $N(1650)S_{11}$. The helicity amplitudes (in units of $10^{-3}$\,GeV$^{-1/2}$) as derived in the fits are listed below.

\begin{center}
\renewcommand{\arraystretch}{1.4}
\begin{tabular}{lcccl}
\hline\hline
  && $N(1535)S_{11}$ & $N(1650)S_{11}$ &\\\hline
$p$ &&   $0.105\pm 0.010$ &   $0.033\pm 0.007$ & \\
$n$ &&   -$0.095\pm 0.006$ &   $0.019\pm 0.006$ &\\
\hline\hline
\end{tabular}
\end{center}

Summarizing, we have shown that the narrow structure observed in the new 
data on photoproduction of $\eta$ mesons off neutrons does not support the
existence of a nucleon resonance with exotic properties.

We would like to thank the A2 Collaboration for providing their data in numerical form.

\end{document}